# *BigFoot:* Analysis, monitoring, tracking and sharing of bio-medical features of human appendages using consumer-grade home and office based imaging devices

Sam Mavandadi[1,2], Steve Feng[1,2], Frank Yu[1,2], Richard Yu[1,2] and Aydogan Ozcan[1,2,3*]

**1** Electrical Engineering Department, University of California, Los Angeles, CA, 90095, USA
**2** Bioengineering Department, University of California, Los Angeles, CA, 90095, USA
**3** California NanoSystems Institute, University of California, Los Angeles, CA, 90095, USA
\* E-mail: ozcan@ucla.edu

http://innovate.ee.ucla.edu/   ;   http://bigfoot.ee.ucla.edu/ ;   http://biogames.ee.ucla.edu/

## Abstract

Here we describe a system for personal and professional management and analysis of bio-medical images captured using off-the-shelf, consumer-grade imaging devices such as scanners, digital cameras, cellphones, webcams and tablet PCs. Specifically, we describe an implementation of this system for the analysis, monitoring and tracking of conditions and features of human feet using a flatbed scanner as the image capture device and a custom-designed set of algorithms and software to manage and analyze the acquired data.

## Background and Motivation

There are various medical conditions that manifest themselves either directly or indirectly as visible features on the human body, e.g., on the feet or arms. As such, effective monitoring and tracking of such superficial features can be of utmost importance as an indirect (and sometimes direct) method of tracking the progression of the underlying medical condition. As an example, diabetes, especially in elderly individuals, can lead to significant visible damage on the surface of e.g., the feet.[1-4] Furthermore, individuals experiencing such conditions are not always aware of the extent and even the existence of the damage and on most occasions they find out about it during visits to the doctor's office or a point-of-care clinic. Given that such trips may not always occur with sufficient frequencies, it might be the case that significant damage can go undetected and untreated for long periods of time. Therefore, a cost-effective and easy-to-use method for the monitoring, tracking and sharing of the condition of the feet (or other human appendages) at home or point-of-care offices is extremely desirable.

## System Overview

Our system hardware consists of an imaging device (e.g. flatbed scanner, digital camera, webcam, etc.), connected to a controller (e.g., local PC, laptop, tablet etc.). The controller may subsequently be connected to a data network (wired or wireless) or an off-site server. Alternatively it can serve as the end-point analytics component (see Figures 1 and 2). The end-point analytics component (e.g. a local PC) will be running a custom-developed analytics software, containing both the controller mechanism for the image capture unit (e.g., a scanner, digital camera, webcam, cellphone etc.), in addition to image processing algorithms that can detect, analyze, track and share the visual features captured by the imaging capture device. Additionally, a data-management system is also included in the system such that the collected images and related information (e.g., text) may be stored and managed off-site on a computer server.

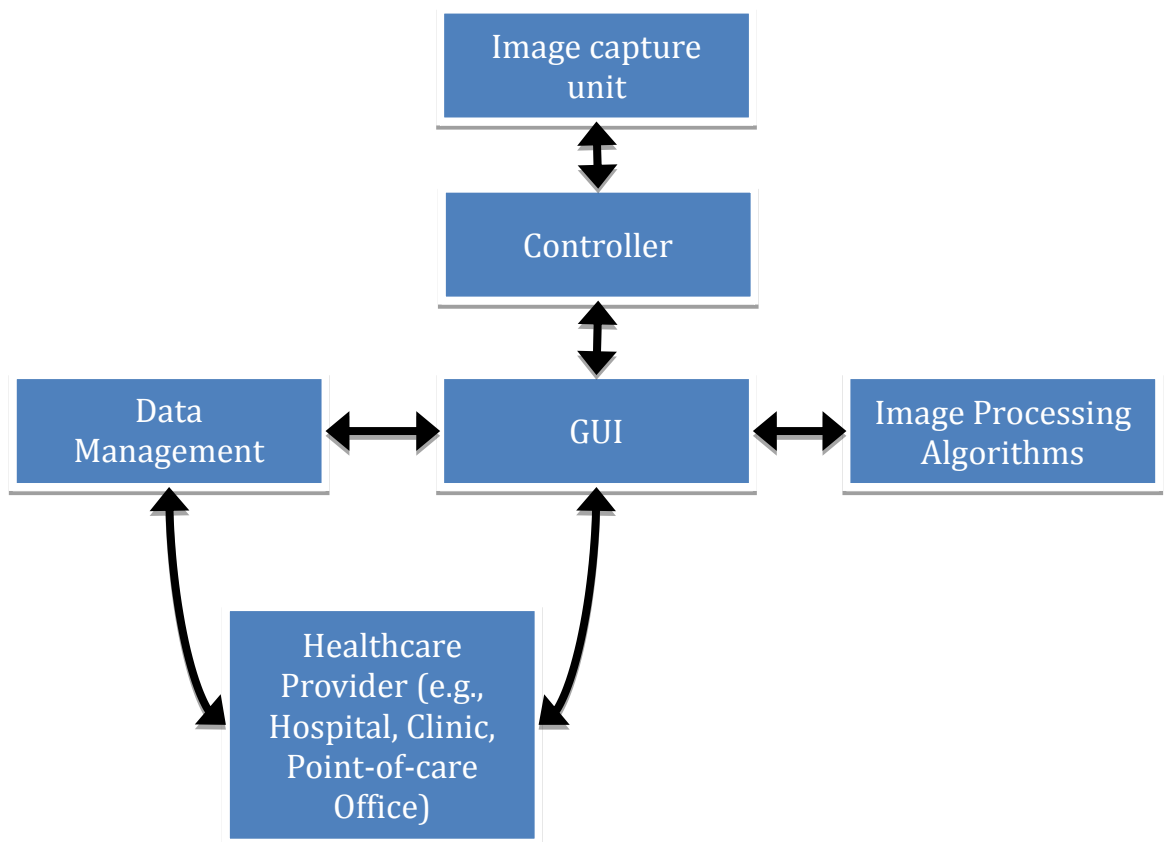

**Figure 1 Overview of the system.** The controller here can be a PC, a laptop, a tablet, or in general any device that can connect to an image capture unit (e.g., a scanner, digital camera, webcam, cellphone, tablet PC etc.) and request images. The Graphical User Interface (GUI) is a platform-specific software designed to assist the user in capturing images and managing the data. A set of custom-developed image processing algorithms are used to analyze, manage, and track the features of interest (e.g., register the images, detect anomalies, track regions of interest, etc.). The system can allow the data to be stored and processed off-site on a data management server. Furthermore, the data can be transmitted on demand or periodically to authorized healthcare providers. There is also a mechanism in place for the authorized healthcare provider to access patient data either directly or from the data management server for more efficient and better care (subject to approval of patients).

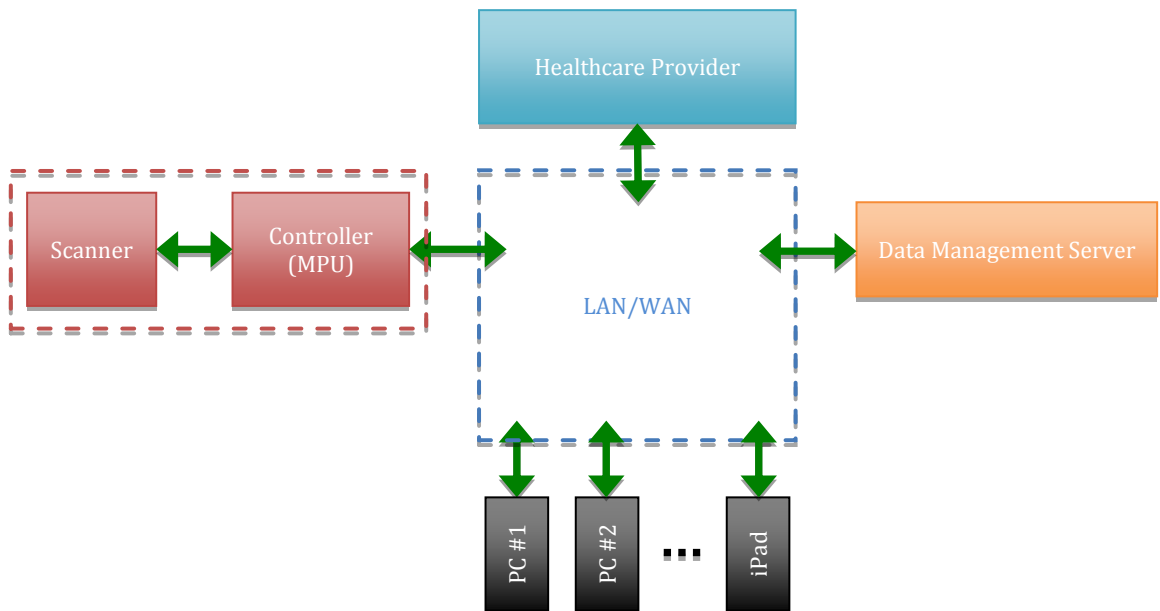

**Figure 2 A broad connection scheme for the system.** In this scheme, the image capture device (for example a scanner) has a dedicated controller connected to it (e.g., a dedicated Linux device). Running on the controller is a dedicated server that in addition to directly controlling the scanner, allows for other devices (e.g., PCs, tablets, smartphones etc.) to connect to the system and request the capturing of bio-medical images. In a multi-user setting (e.g., hospital), the controller is connected to a Local Area Network (LAN) or a Wireless Area Network (WAN) and can accept connections from authorized devices connected to this Intranet. The data is processed and stored on a data management server and is made available to authorized medical personnel. Note that in this scheme, all of the processing is done on the data management server, and the viewing devices (e.g., PCs, tablets, etc.) are not required to do any computation.

## Subsystems

### Image Capture Device

The Image Capture Device can be any electronic device capable of capturing biomedical images of interest (e.g., a scanner, digital camera, webcam, cellphone, tablet PC, etc.). Without loss of generality, in our implementation, we have made use of a consumer-grade flatbed scanner to capture images of human feet. The same platform can be generalized to capture images from any bodily surface (e.g., palm, chest, arms etc.).

### Controller

In general, the Controller subsystem can be any device capable of making connection calls to the Image Capture Device(s), requesting and receiving image data. The most basic controller subsystem is a personal computer or a tablet PC, connected to the scanner. This might be one of the most typical scenarios for home and office use, where the PC/tablet (which might also serve as the analysis and viewing device) is connected to the scanner through e.g., a USB or Wi-Fi connection (see Figure 1).

In a more general setting, the controller subsystem can be completely independent of the viewing, storage, and analysis devices (see Figure 2). A typical controller can be a very basic single-board computer running an operating system such as Linux or Android. This controller is connected to the scanner through some sort of data connection (e.g., USB) and also connected to some sort of LAN/WAN. It will be running a custom-designed server which has the job of responding to incoming requests through the LAN/WAN for initiating image capture jobs; following a request, the data is transmitted either to the requesting party, or to a data management server, depending on the configuration of the system. As an example, a patient making a scan of his feet can make a request to the controller through a custom-designed application running on e.g., his smartphone (both the smartphone and the controller are connected to a LAN/WAN). The controller instructs the scanner to initiate the scan, following which the image is transmitted to the data management server indicated on the patient's system configuration. The data is then subsequently analyzed off-site and a resized image and the analysis report are then transmitted and displayed back on the patient's smartphone. Depending on the configuration, the scan data may be automatically (or by request of the patient) transmitted to his/her authorized physician and incorporated into patient's medical records.

Note that in a setting such as a hospital or a point-of-care office, the controller can receive requests from multiple authorized devices. For example, multiple authorized physicians can make use of the same imaging system to scan their patients using their own individual PCs/Tablets/Smartphones.

## Data Management Server

This is a custom-designed server for the purposes of storing, management and potentially analysis, of the captured image dataset. Depending on the type of use (e.g., physicians, patients, hospitals, etc.), the users may elect to store and analyze their data off-site. The system can also be configured so that (conditioned on the availability of an internet connection) the data can be directly transmitted to such a server. Furthermore, the users can elect between storage-only and store-and-process schemes.

## Image Processing Algorithms

These are a set of extendable algorithms for the analysis of the captured images. In the case of feet imaging, some exemplary algorithms include orientation detection, registration, scar detection and tracking, segmentation, region of interest (ROI) correspondence detection, etc. The software is designed in such a way that new algorithms can be added as plugins, as needed.

## Software Graphical User Interface (GUI)

The main [Graphical User Interface (GUI)](), running on a personal computer, tablet or a smartphone allows the user to request scans and manage, analyze, track and share the captured images. Figures 3-9 show screenshots of our first prototype for this GUI.

In the specific example of feet imaging, following each scan, the software automatically detects the orientation of the foot, and registers the image to prior images that were taken for the same patient. Therefore, the spatial correspondence between the different images is always known, making the tracking of ROIs much easier.

A user is allowed to select ROIs on the images. Given that the spatial correspondences between the images are known, once an ROI is marked in one image, it can be easily tracked over all other available images, allowing one to monitor the condition and time-evolution of locations of interest (e.g., scars, skin anomalies, etc.).

Through the use of image processing algorithms, a user can also automatically detect scars and regions of interest on an image, mark them, and then track their evolution forward or backward in time. The user can also enter text that can be attached to each region of interest. This text can be edited by authorized users such as the patient or his/her physician(s).Regions of interest, associated text, and other derived information can then be transmitted to the data management server or authorized healthcare providers for off-site storage and/or further analysis.

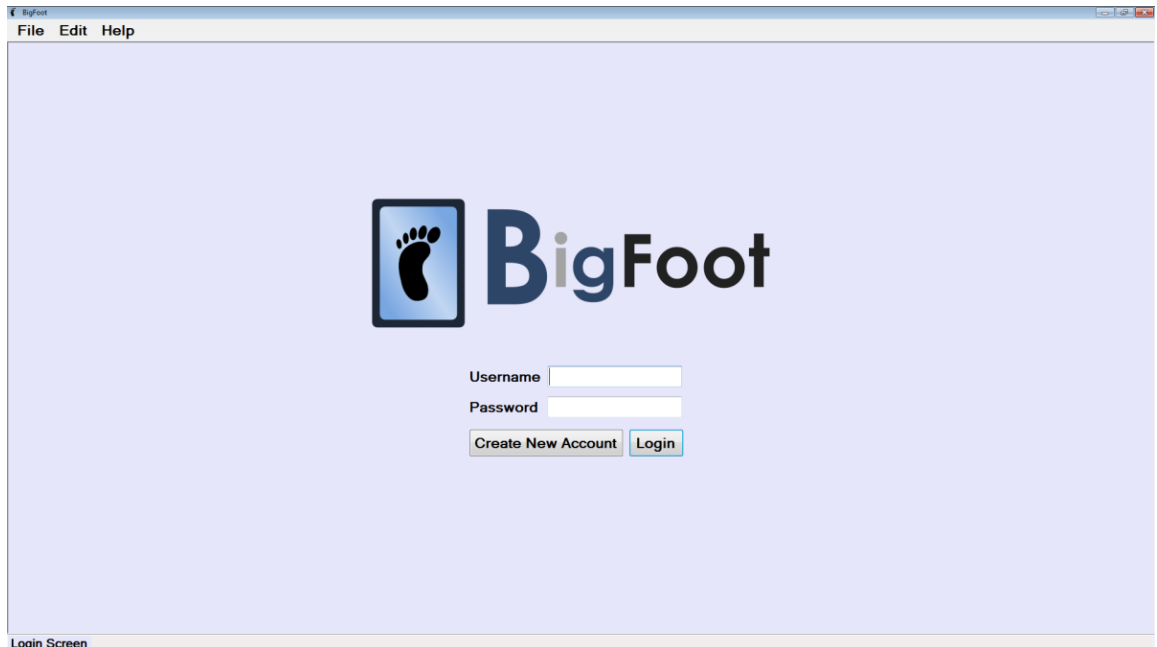
**Figure 3 User log-in window.** To download the latest application: http://bigfoot.ee.ucla.edu/

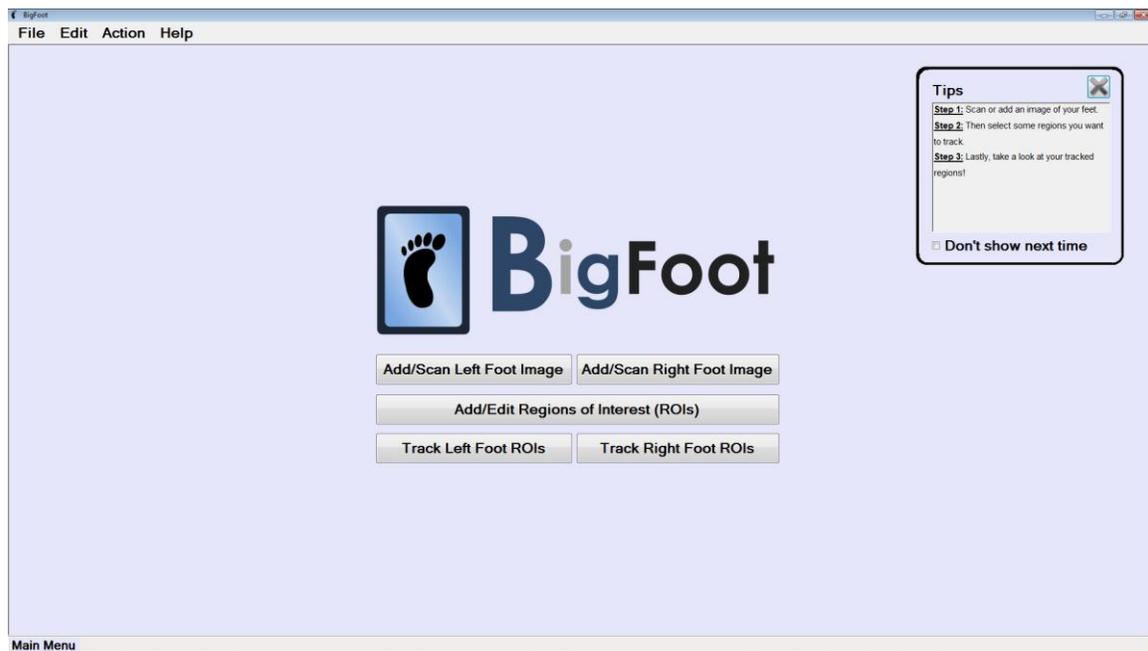
**Figure 4 Main window.** The user can choose between taking new images, selecting ROIs, or tracking previously designated ROIs. Various popup windows provide helpful instructions on using this specific application. To download the latest application: http://bigfoot.ee.ucla.edu/

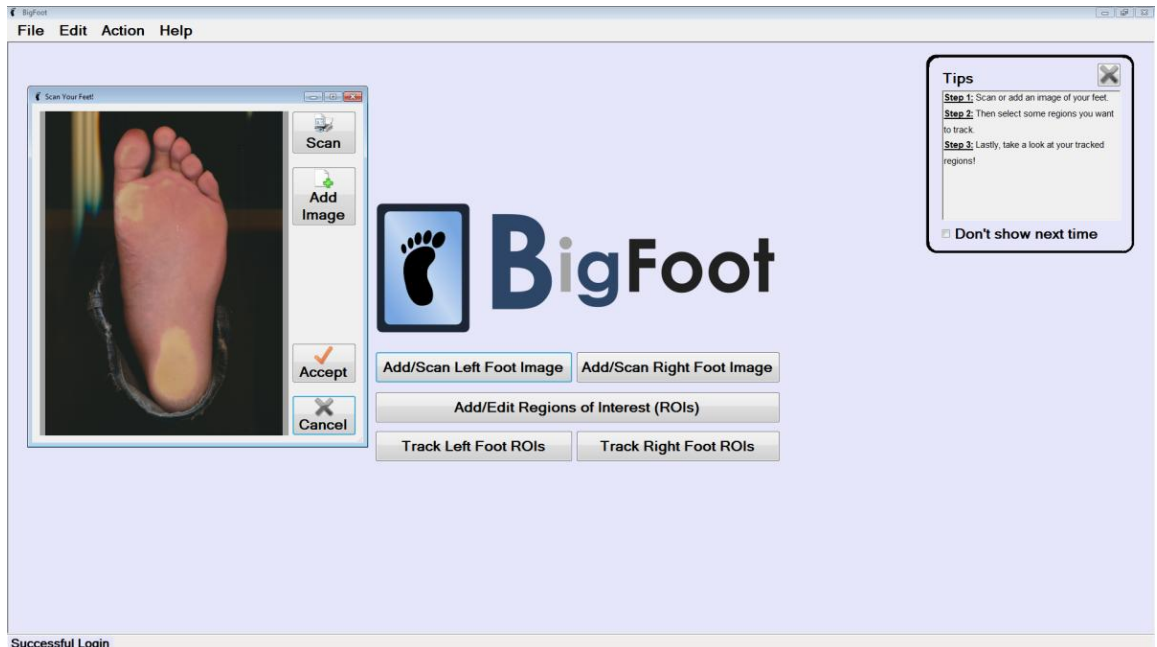
**Figure 5 Scan window.** In this case the user has already scanned his right foot. To download the latest application: http://bigfoot.ee.ucla.edu/

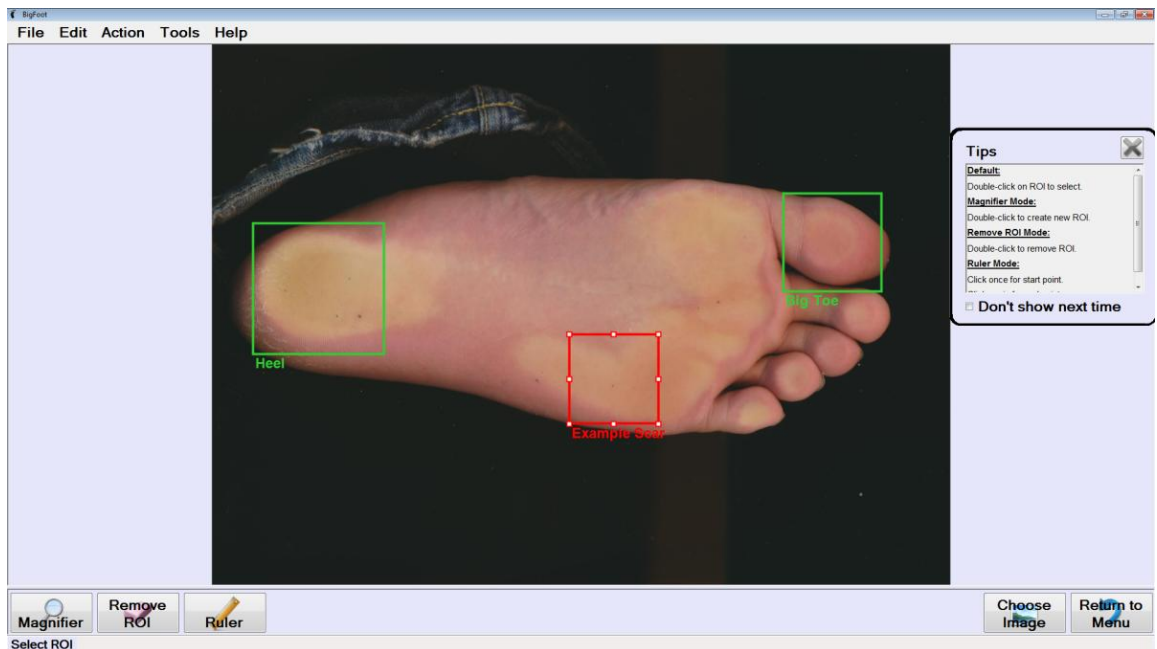
**Figure 6 ROI selection window.** The user has selected ROIs on a scanned image. The red selected ROI can be freely resized, moved, and deleted. The green boxes indicate already selected and approved ROIs. To download the latest application: http://bigfoot.ee.ucla.edu/

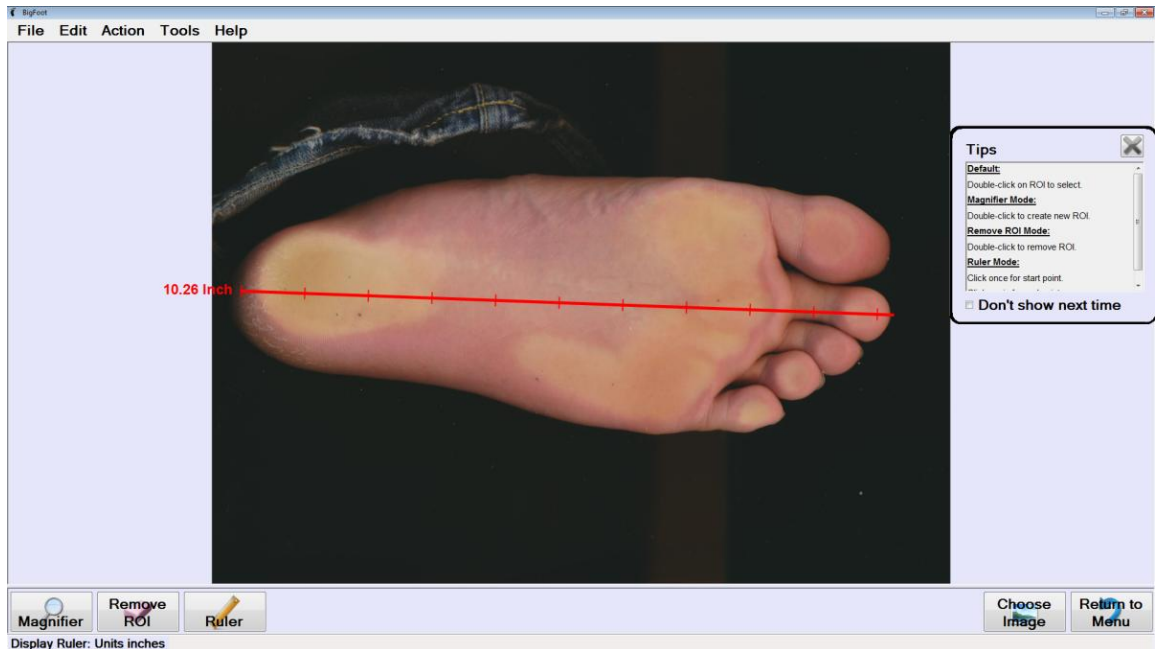
**Figure 7 Image analysis window.** The user can visually measure portions of the feet. Other analytical tools can be added as plugins, and will appear as buttons on the bottom tool bar. To download the latest application: http://bigfoot.ee.ucla.edu/

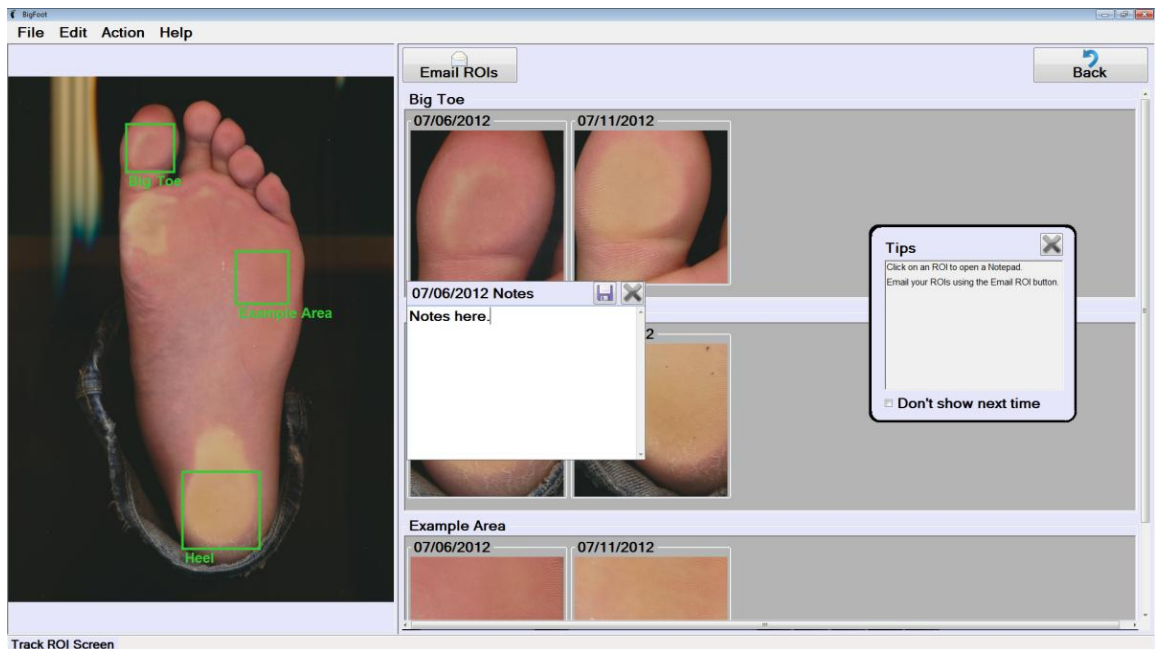
**Figure 8 ROI Tracking window.** The user can view the time sequence of selected ROIs (automatically detected following registration) as imaged on different dates. A notepad allows tagging notes (by e.g., the patient or other authorized users) on each ROI individually. To download the latest application: http://bigfoot.ee.ucla.edu/

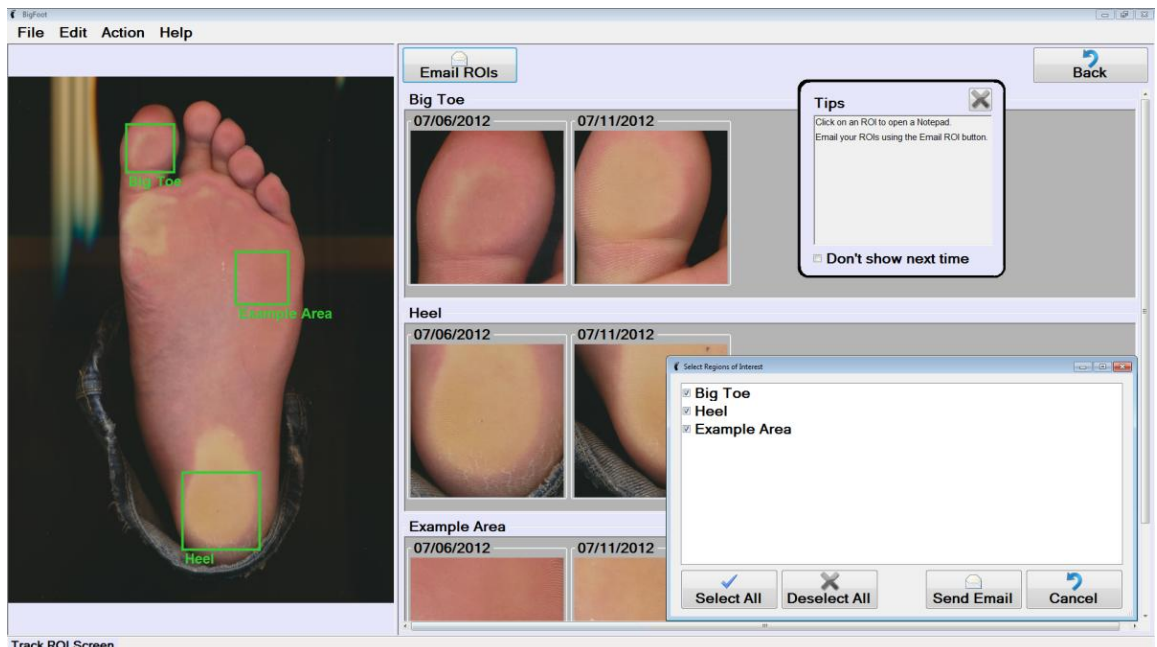

**Figure 9 ROI window.** The user has the ability to email the specific ROI information or a sequence of ROIs to his/her physician along with text that can be tagged for each ROI. The text also serves as a log of the communication and information exchange between the patient and the medical expert(s). To download the latest application: http://bigfoot.ee.ucla.edu/